\documentclass[preprint]{aastex62}
\usepackage{graphicx}

\def\kms{$\textrm{km~s$^{-1}$}$}

\def\H2{H$_{2}$}

\def\roH2{$\rho_{\textrm{H}_2}$}
     
\def\MH2{M$_{\textrm{H}_2}$}

\newcommand\msun{\rm\,M\sun}
\newcommand\lsun{\rm\,L\sun}

\newcommand\myr{\msun \, {\rm yr}^{-1}}
\newcommand{\MC}{\multicolumn}
\newcommand\hii{H\,{\sc ii} \,}

\def\apgt{\ {\raise-.5ex\hbox{$\buildrel>\over\sim$}}\ }
\def\aplt{\ {\raise-.5ex\hbox{$\buildrel<\over\sim$}}\ }

\submitjournal{AJ}

\shorttitle{Two circumstellar nebulae and their central stars} \shortauthors{Gvaramadze et al.}

\begin{document}

\title{Two circumstellar nebulae discovered with the {\it Wide-field Infrared Survey Explorer} and their massive central stars}

\correspondingauthor{Vasilii Gvaramadze}

\author{Vasilii V. Gvaramadze}
\affil{Sternberg Astronomical Institute, M.V. Lomonosov Moscow State University, Universitetsky pr., 13, Moscow, 119991 Russia}
\affil{Space Research Institute, Russian Academy of Sciences, Profsoyuznaya 84/32, 117997 Moscow, Russia}
\affil{Isaac Newton Institute of Chile, Moscow Branch, Universitetskij Pr. 13, Moscow 119992, Russia}
\email{vgvaram@mx.iki.rssi.ru}

\author{Alexei Yu. Kniazev}
\affil{South African Astronomical Observatory, PO Box 9, 7935 Observatory, Cape Town, South Africa}
\affil{Southern African Large Telescope Foundation, PO Box 9, 7935 Observatory, Cape Town, South Africa}
\affil{Sternberg Astronomical Institute, M.V. Lomonosov Moscow State University, Universitetsky pr., 13, Moscow, 119991 Russia}

\author{Norberto~Castro}
\affil{Department of Astronomy, University of Michigan, 1085 S. University Avenue, Ann Arbor, MI 48109-1107, USA}

\author{Eva K.~Grebel}
\affil{Astronomisches Rechen-Institut, Zentrum f\"ur Astronomie der Universit\"at Heidelberg, M\"onchhofstr. 12-14, D-69120}

\begin{abstract}
We report the discovery of two mid-infrared nebulae in the northern hemisphere with the {\it Wide-field Infrared Survey Explorer} 
and the results of optical spectroscopy of their central stars, BD+60$\degr$\,2668 (composed of two components, separated from 
each other by $\approx$3 arcsec) and ALS\,19653, with the Calar Alto 3.5-m telescope and the Southern African Large Telescope 
(SALT), respectively. We classify the components of BD+60$\degr$\,2668 as stars of spectral types B0.5\,II and B1.5\,III. 
ALS\,19653 is indicated in the SIMBAD data base as a planetary nebula, while our observations show that it is a massive B0.5\,Ib 
star, possibly in a binary system. Using the stellar atmosphere code {\sc fastwind}, we derived fundamental parameters of the 
three stars as well as their surface element abundances, implying that all of them are either on the main sequence or only 
recently left it. This provides further evidence that massive stars can produce circumstellar nebulae while they are still
relatively unevolved. We also report the detection of optical counterparts to the mid-infrared nebulae and a second, more extended
optical nebula around ALS\,19653, and present the results of SALT spectroscopy of both nebulae associated with this star. The 
possible origin of the nebulae is discussed.
\end{abstract}


\section{Introduction}
\label{sec:int}

Mid-infrared (IR) surveys of the inner regions of the Galactic plane and several other sites of massive star formation in the 
Milky Way carried out by the {\it Spitzer Space Telescope} led to the discovery of many dozens of nebulae reminiscent of 
circumstellar nebulae around evolved massive stars (Gvaramadze, Kniazev \& Fabrika 2010b; Wachter et al. 2010; Mizuno et al. 2010). 
Follow-up spectroscopic observations of central sources of these nebulae showed that many of them are massive stars at
different evolutionary stages (e.g. Gvaramadze et al. 2009, 2010b; Wachter et al. 2010; Stringfellow et al. 2012a, 2012b; Flagey 
et al. 2014; Kniazev \& Gvaramadze 2015; Silva et al. 2017). Although {\it Spitzer} has covered the areas of the sky where most 
IR nebulae produced by massive stars are expected to reside, a significant number of yet undetected nebulae might remain 
to be uncovered. With the advent of the {\it Wide-field Infrared Survey Explorer} ({\it WISE}; Wright et al. 2010) and its all-sky 
survey it became possible to search for mid-IR nebulae in regions not observed by {\it Spitzer}, which allowed us to discover 
several dozens of new nebulae (e.g. Gvaramadze et al. 2012).

The origin of circumstellar nebulae around massive stars is primarily due to copious mass loss inherent to these stars at different 
stages of their life. For example, the compact (pc-scale) nebulae around Wolf-Rayet stars of late nitrogen sequence are 
produced in the course of interactions between the fast Wolf-Rayet wind and the slow, dense material lost during the preceding red 
supergiant phase (e.g. Brighenti \& D'Ercole 1995; Garcia-Segura, Langer \& Mac Low 1996). The red supergiants themselves can also 
produce compact circumstellar nebulae, provided that their wind is confined by some external factors (e.g. Morris \& Jura 1983; 
Mackey et al. 2014). Most often, the circumstellar nebulae are found around (candidate) luminous blue variables
(LBVs) (Nota et al. 1995; Clark, Larionov \& Arkharov 2005), and it is believed that they are produced by either instant mass 
ejections or by brief (maybe recurrent) episodes of enhanced mass loss (Humphreys \& Davidson 1994). Currently more than 70 per cent 
of stars of this type are known to be associated with nebulae of various shapes (Kniazev, Gvaramadze \& Berdnikov 2015).

The bipolar morphology of some nebulae associated with massive stars suggests that their formation is somehow related to the (nearly 
critical) rotation of their underlying stars (Langer 1998). The high rotational velocity of massive stars could be intrinsic 
to some of them from birth or could be achieved in the course of stellar evolution, e.g. because of contraction of a single star 
during the transition from core-H to core-He burning or because of mass transfer in a close binary system. Since the majority of 
massive stars form in binary or multiple systems (e.g. Chini et al. 2012; Sana et al. 2012), the binarity may actually be the most 
important factor responsible for the origin of circumstellar nebulae. Indeed, the presence of a second star 
may not only spin-up its companion, but also may trigger various modes of mass loss from the system, as it was proposed 
to explain a variety of shapes of planetary nebulae (e.g. Morris 1981; Livio, Salzman \& Shaviv 1979; Fabian \& Hansen 1979; 
Corradi \& Schwarz 1993; Mastrodemos \& Morris 1999; De Marco 1999). Moreover, a significant 
fraction of massive stars are subject to binary interaction processes (including mergers) already during the main sequence stage 
(Sana et al. 2012; de Mink et al. 2014), implying that circumstellar nebulae could be produced by relatively unevolved stars.

The stellar material ejected in the course of binary interaction, e.g. during the common envelope evolution or merger of the 
binary components, is expected to be concentrated close to the orbital plane of the system. The resulting flattened (disc-like)
circumstellar structure can, in principle, collimate the fast wind of the post-interaction binary or the merger product of two stars
in the polar directions. A strong magnetic field generated during the common envelope stage (Reg\"os \& Tout 1995; Tout \& Reg\"os 
2003) or a merger of the binary components (Langer 2012; Wickramasinghe, Tout \& Ferrario 2014) could affect the 
geometry of the resulting circumstellar nebulae as well (cf. Chevalier \& Luo 1994; R\'o\.zyczka \& Franco 1996; Garcia-Segura et 
al. 1999; Nordhaus \& Blackman 2006).

Another potentially important mechanism for the origin of circumstellar nebulae around single (or apparently single) massive 
stars is connected to the bi-stability jump (Pauldrach \& Puls 1990; Lamers \& Pauldrach 1991) --- an abrupt (a factor of 10) 
increase in the wind mass-loss rate (Vink 2018), $\dot{M}$, when the stellar effective temperature, $T_{\rm eff}$, 
decreases below the critical value of $\sim21$\,kK (Lamers, Snow \& Lindholm 1995), which corresponds 
to a spectral type of around B1. The increase in $\dot{M}$ is accompanied (Lamers et al. 1995) by a factor of 2 decrease in the 
terminal wind velocity, $v_\infty$, leading to a factor of $\sim20$ increase in the density of the stellar wind. In fast-rotating 
stars the bi-stability jump may occur preferentially at low stellar latitudes because of the effect of gravity darkening, which could 
result in an equatorial outflow (Lamers \& Pauldrach 1991; Lamers et al. 1995).

Lamers et al. (2001) analysed the chemical composition of nebulae around several LBVs and came to the conclusion that they were ejected 
soon after their central stars left the main sequence and that the observed chemical enhancements are due to rotationally induced 
mixing during the main sequence stage. Subsequent studies of blue supergiants with LBV-like circumstellar nebulae (Smartt et al. 2002; 
Hendry et al. 2008; Smith et al. 2013; Gvaramadze et al. 2014, 2015, 2018; Mahy et al. 2016) lent further support to the idea that 
massive stars can form circumstellar nebulae during early phases of their evolution. Two more such stars are presented in this paper. 

In Section\,\ref{sec:two}, we show multi-wavelength images of two nebulae discovered with {\it WISE} and review the existing 
information on their central stars. Section\,\ref{sec:obs} describes our optical spectroscopic observations (preliminary results of 
these observations were reported in Gvaramadze \& Kniazev 2017). The spectral analysis of the stars is given in Section\,\ref{sec:ana}. 
In Section\,\ref{sec:neb}, we present and discuss spectra of an optical counterpart to one of the mid-IR nebulae. Our results are
discussed in Section\,\ref{sec:dis} and summarized in Section\,\ref{sec:sum}. 

\section{Two {\it WISE} nebulae and their central stars}
\label{sec:two}

\begin{figure}
\includegraphics[width=8cm]{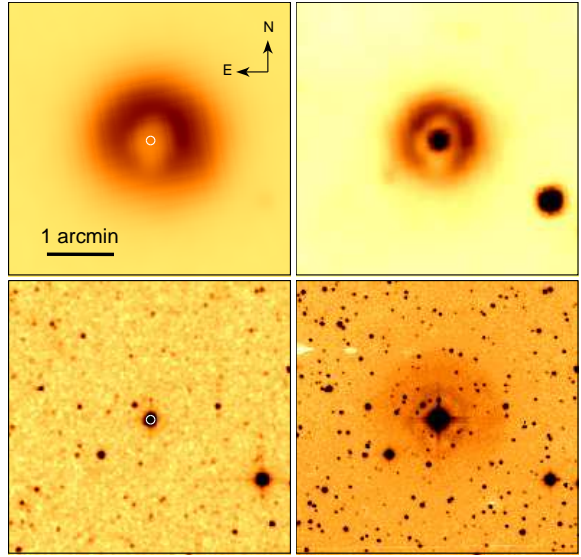}
\centering \caption{From left to right, and from top to bottom: {\it WISE} 22 and 12\,\micron, 2MASS $K_{\rm s}$-band and IPHAS
H$\alpha$ images of the region containing BD+60$\degr$\,2668 (indicated by a circle) and its circumstellar shell WS3. The orientation
and the scale of the images are the same. At a distance of 3.6 kpc, 1 arcmin corresponds to $\approx1$ pc. Note the 
presence of two sets of (horizontal) diffraction spikes in the IPHAS image, indicating that BD+60$\degr$\,2668 is composed of two 
components separated by $\approx3$ arcsec (see text for details).}
\label{fig:ws3}
\end{figure}

\begin{figure*}
\includegraphics[width=12cm]{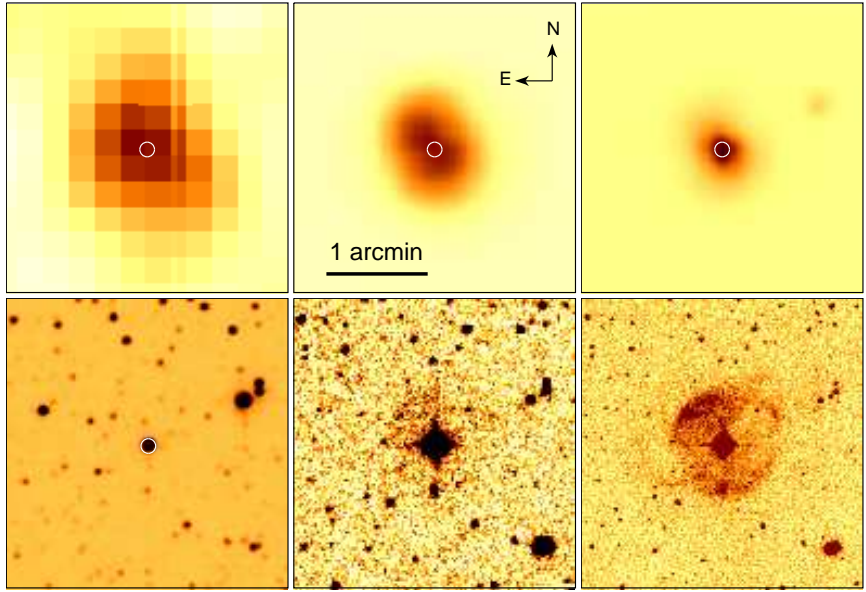}
\centering \caption{From left to right, and from top to bottom: NVSS 1.4 GHz, {\it WISE} 22 and 12\,\micron, 2MASS $K_{\rm s}$-band,
DSS-II red band and IPHAS H$\alpha$ images of the region containing ALS\,19653 (indicated by a circle) and its circumstellar shell WS29. 
The orientation and the scale of the images are the same. At a distance of 1.57 kpc, 1 arcmin corresponds to $\approx0.45$ pc.} 
\label{fig:ws29}
\end{figure*}

In a search for rare types of massive stars through the detection of their circumstellar nebulae with {\it WISE}, we discovered several 
dozens of nebulae in the regions not covered by {\it Spitzer}. Two of them, named WS3 and WS29 (where WS stands for ``{\it WISE} 
shell"; cf. Gvaramadze et al. 2012), are presented below. These nebulae positionally coincide with the {\it Infrared Astronomical Satellite} 
({\it IRAS}) sources, respectively, IRAS\,00033+6035 and IRAS\,18454+0250. The latter source is indicated in the SIMBAD data 
base\footnote{http://simbad.harvard.edu/simbad/} as a planetary nebula. For both nebulae we found optical counterparts in the 
Isaac Newton Telescope (INT) Photometric H$\alpha$ Survey of the Northern Galactic Plane (IPHAS; Drew et al. 2005),
which are indicated as circumstellar matter in the Hong Kong/AAO/Strasbourg H$\alpha$ (HASH) planetary nebula database (Parker, 
Boji{\v c}i{\'c} \& Frew 2016; see also Kronberger et al. 2016), and named there as PNG\,117.4$-$01.5 (or Pa\,58) 
and PNG\,035.1+02.0, respectively.
WS29 is also discernable in the Digitized Sky Survey II (DSS-II; McLean et al. 2000) red band image and is a known source of radio 
emission (Condon, Kaplan \& Terzian 1999). Figs\,\ref{fig:ws3} and \ref{fig:ws29} present {\it WISE} 22 and 12\,\micron, Two Micron 
All Sky Survey (2MASS; Skrutskie et al. 2006) $K_{\rm s}$-band and IPHAS H$\alpha$ images of the regions containing WS3 and WS29. 
Fig.\,\ref{fig:ws29} also gives the DSS-II and 1.4 GHz National Radio Astronomy Observatory (NRAO) Very Large Array (VLA) Sky Survey 
(NVSS; Condon et al. 1998) images of WS29.

Fig.\,\ref{fig:ws3} shows that WS3 is visible in both {\it WISE} images: at 22\,\micron \, it appears as a thick, diffuse, almost circular 
shell with a radius of $\approx$1 arcmin, while at 12\,\micron \, it appears as a slightly elongated shell with a smaller radius of 
$\approx$35 arcsec and a point-like source in the centre. In both images, the brightness of the shell is reduced to the south. In the 
IPHAS H$\alpha$ image WS3 appears as a slightly elongated shell (of similar angular size as the 12\,\micron \, shell) immersed in a 
diffuse halo of the same extent as the emission at 22\,\micron. The brightness of the halo also decreases to the south, while the inner 
shell appears to be more extended in this direction. The observed brightness asymmetry might be caused by a density gradient in the local interstellar medium or by motion of the local medium or the nebula itself.

Using the SIMBAD data base, we identified the central source of WS3 with the optical star BD+60$\degr$2668, known as a double or multiple 
system (Douglass et al. 1999). A literature search revealed that BD+60$\degr$2668 was recognized as an OB star by Nassau \& Morgan (1951) 
and classified as B2 by Brodskaya (1953), as B1\,III by Morgan, Code \& Whitford (1955), and more recently as B1 by Radoslavova (1989). 
According to Douglass et al. (1999), BD+60$\degr$2668 is composed of two stars, hereafter BD+60$\degr$2668A and BD+60$\degr$2668B, 
separated from each other by $\approx$3 arcsec, and lying on a line with a position angle (PA) of $\approx$159$\degr$ (measured from 
north to east). The second data release of {\it Gaia} (DR2; Gaia Collaboration 2018) provides for BD+60$\degr$\,2668A and 
BD+60$\degr$\,2668B accurate parallaxes of $0.2827\pm0.0290$ mas and $0.2643\pm0.0347$ mas, placing these stars at 
$d=3.54^{+0.40} _{-0.33}$ kpc and $3.78^{+0.57} _{-0.44}$ kpc, respectively. In what follows, we adopt a distance of 3.6 kpc to both 
stars as well as to WS3 (cf. Section\,\ref{sec:ws3}). At this distance, 1 arcmin corresponds to $\approx1$ pc.
Some of the properties of BD+60$\degr$2668A and BD+60$\degr$2668B are summarized in Table\,\ref{tab:det}. 

\begin{table}
  \caption{Properties of BD+60$\degr$\,2668A\&B and ALS\,19653. The spectral types are based on our spectroscopic observations. The 
  coordinates and photometry of BD+60$\degr$\,2668A\&B are from Kharchenko (2001), while those of ALS\,19653 are from Henden 
  et al. (2016).}
  \label{tab:det}
  \begin{center}
  \begin{tabular}{lccc}
      \hline
      & BD+60$\degr$\,2668A & BD+60$\degr$\,2668B & ALS\,19653 \\
      & (WS3)               & (WS3)               & (WS29)     \\
      \hline
      SpT & B0.5\,II & B1.5\,III & B0.5\,Ib \\
      $\alpha$ (J2000) & $00^{\rm h} 06^{\rm m} 01\fs39$ & $00^{\rm h} 06^{\rm m} 01\fs52$ & $18^{\rm h} 48^{\rm m} 00\fs66$ \\
      $\delta$ (J2000) & $60\degr 52\arcmin 02\farcs0$ & $60\degr 51\arcmin 59\farcs1$ & $02\degr 54\arcmin 17\farcs1$ \\
      $l$ & 117\fdg4303 & 117\fdg4304 & 35\fdg1279 \\
      $b$ & $-$1\fdg5185 & $-$1\fdg5193 & 2\fdg0859 \\
      $B$ (mag) & $9.86\pm0.02$ & $10.33\pm0.03$ & $14.40\pm0.04$ \\
      $V$ (mag) & $9.48\pm0.02$ & $10.00\pm0.04$ & $12.40\pm0.05$ \\
      \hline
    \end{tabular}
\end{center}
\end{table}

\begin{figure}
\includegraphics[width=10cm]{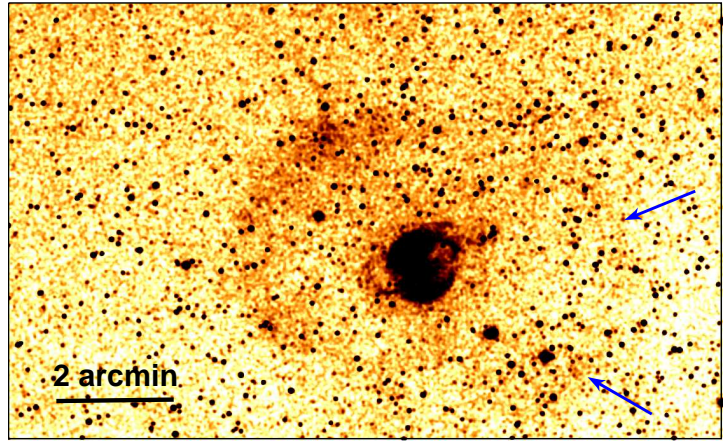}
\centering \caption{IPHAS H$\alpha$ image of a region around WS29 showing the presence of a second, more extended shell. 
This shell is more prominent to the east and northeast, but can also be discerned to the west and southwest from WS29 
(as indicated by arrows). North is up, east is left.} 
\label{fig:ws29-2}
\end{figure}

The IR morphology of WS29 differs completely from that of WS3 (see Fig.\,\ref{fig:ws29}). At 22\,\micron \, WS29 has a dumbbell-like 
shape with the long axis oriented in the northeast-southwest direction (PA$\approx$30$\degr$). At 12\,\micron \, the nebula appears as 
a centrally-brightened source, slightly elongated in the same direction as the 22\,\micron \, nebula. In the H$\alpha$ image WS29 appears 
as an almost circular shell of angular diameter of about 1 arcmin with enhanced brightness on the northeast and southwest sides,
i.e. in the places where the IR nebula is in apparent contact with the optical shell. This image also shows several filaments extending 
in the northwest and southeast directions beyond the main shell, suggesting that there are protrusions in the shell along the line 
perpendicular to the IR nebula (PA$\approx120\degr$). The IPHAS H$\alpha$ image also shows that WS29 is surrounded by a more 
extended shell of angular radius of about 2.7 arcmin. This shell is more prominent to the east and northeast, but can also
be discerned to the west and southwest of WS29 (see Fig.\,\ref{fig:ws29-2} and Sections\,\ref{sec:neb} and \ref{sec:ws29}).

Using the SIMBAD data base, we found that the central star of WS29, known as ALS\,19653 (or PDS\,543), was classified as a B1 star by 
Vieira et al. (2003). Condon et al. (1999) detected an elongated radio source at the position of WS29 in the 1.4 GHz NVSS image (see 
Fig.\,\ref{fig:ws29}) with the major and minor axes of $\approx$77 and 47 arcsec, respectively. The PA of the major axis of 
$\approx24\degr$ is similar to that of the WISE 22\,\micron \, nebula. Moreover, polarimetric observations of ALS\,19653 by Rodrigues et 
al. (2009) revealed an intrinsic polarization of the star of $1.14\pm0.04$ per cent with a polarization angle of 121$\degr$. This 
implies the presence of a non-spherical structure around the star with the major axis at PA$\approx$30$\degr$ (i.e. parallel to the 
major axis of the 22\,\micron \, nebula). The {\it Gaia} DR2 parallax of ALS\,19653 of $0.6388\pm0.0633$ mas (Gaia Collaboration 2018) 
places this star at $d=1.57^{+0.17} _{-0.14}$ kpc. At a distance of 1.57 kpc, 1 arcmin corresponds to $\approx0.45$ pc. Some of the 
properties of ALS\,19653 are summarized in Table\,\ref{tab:det}. Note that ALS\,19653 is much more reddened compared to 
BD+60$\degr$\,2668A\&B ($B-V=2$ mag versus $B-V\approx0.3$ mag) despite a factor of two shorter distance to the former star.

\section{Spectroscopic observations}
\label{sec:obs}

\subsection{BD+60$\degr$\,2668A\&B}
\label{sec:bd}

Spectra of BD+60$\degr$\,2668A\&B were obtained with the TWIN spectrograph attached to the Cassegrain focus of the 3.5-m telescope in 
the Observatory of Calar Alto (Spain) on 2012 July 12. Three exposures of 60\,s were taken. The set-up used for TWIN consisted of 
the grating T08 in the first order for the blue arm (spectral range of 3500--5600 \AA) and T04 in the first order for the red arm 
(spectral range of 5300--7600 \AA), which provides a reciprocal dispersion of 72~\AA\,${\rm mm}^{-1}$ for both arms. The resulting 
full width at half-maximum (FWHM) spectral resolution measured on strong lines of the night sky and reference spectra was 3.1--3.7~\AA. 
The slit of $240\times2.1$ {\rm arcsec}$^2$ was oriented at a PA selected in a way to observe both stars simultaneously. The seeing 
during the observations was not stable, $\simeq1.0-1.5$ arcsec. Spectra of He--Ar comparison arcs were obtained to calibrate the 
wavelength scale and spectrophotometric standard star BD+$33\degr$\,2642 (Bohlin 1996) was observed at the beginning of 
the night for the flux calibration. 

\begin{figure*}
\begin{center}
\includegraphics[width=12cm,angle=270,clip=]{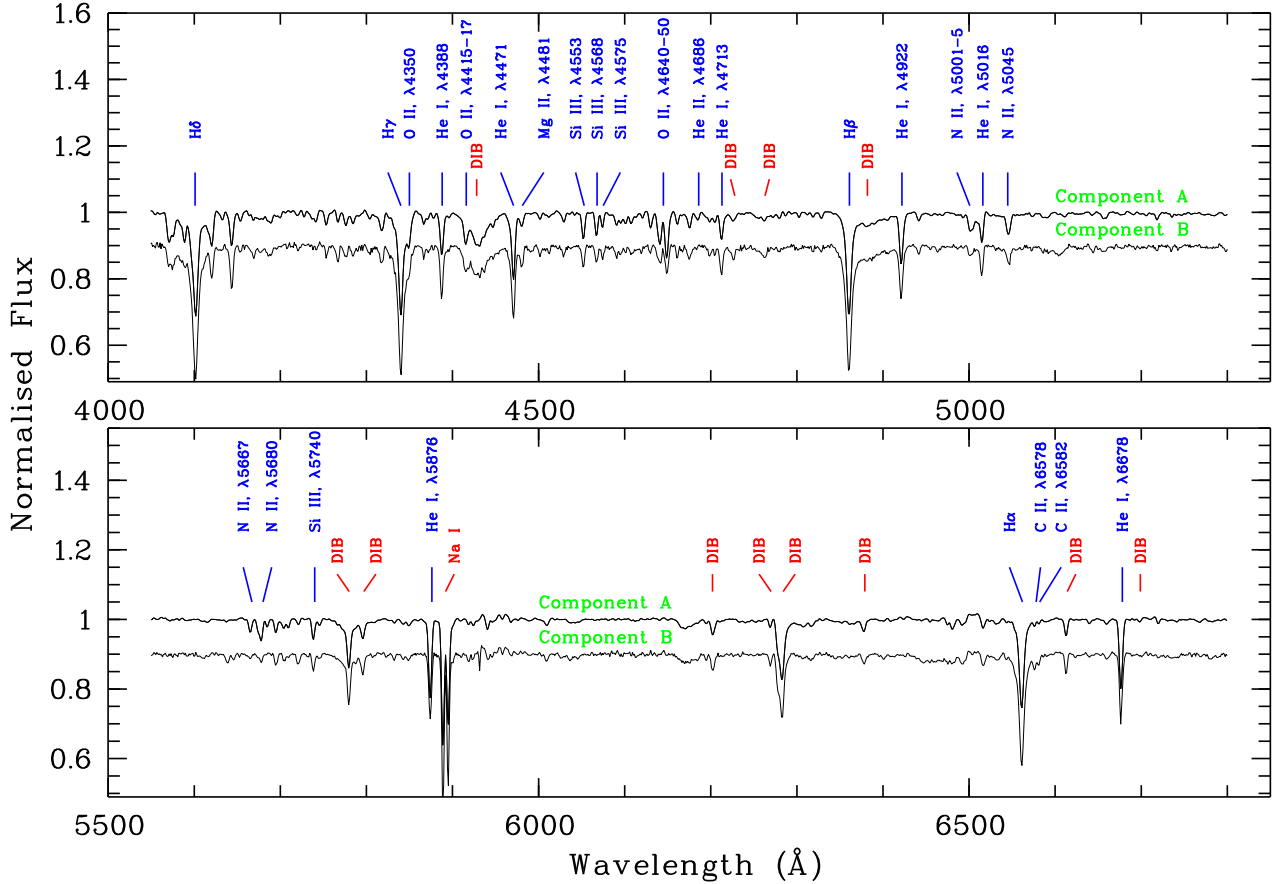}
\end{center}
\caption{Normalized spectra of two components of BD+60$\degr$\,2668 obtained with the TWIN spectrograph at the 3.5-m telescope at the 
Calar Alto Observatory. The principal lines and most prominent diffuse interstellar bands (DIBs) are indicated.}
\label{fig:bd}
\end{figure*}

The primary data reduction was done using the {\sc iraf} package. The data for each CCD detector were trimmed, bias subtracted, and flat
corrected. The subsequent long-slit data reduction was carried out in the way described in Kniazev et al. (2008). The two-dimensional 
(2D) spectra were averaged and one-dimensional (1D) spectra for both components were then extracted using the {\sc iraf} {\sc apall} 
task. The resulting normalized 1D spectra are shown in Fig.\,\ref{fig:bd}.

Unfortunately, the optical counterpart to WS3 was discovered after the spectra of the central stars were obtained. The short total 
exposure used for spectroscopy of these bright stars did not allow us to detect signatures of the nebular emission in the 2D spectrum. 

\subsection{ALS\,19653}
\label{sec:als}

\begin{table*}
\caption{Journal of SALT observations of ALS\,19653 and its circumstellar nebula WS29.}
\label{tab:obs}
\begin{tabular}{llccccccc} \hline
Instrument & Date & Exposure & Spectral scale & Slit & PA & Seeing & Spectral Range \\
 & & (sec) & (\AA\,pixel$^{-1}$) & (arcsec) & ($\degr$) & (arcsec) & (\AA) \\
 \hline
RSS & 2016 April 25  & 240  & 0.97  & 1.25 & 0   & 1.6 & 4200$-$7300 \\
HRS & 2016 May 27  & 1500 & 0.04  & --   & --  & 3.0 & 3700$-$8900 \\
RSS & 2016 May 28  & 1200 & 0.26  & 2.00 & 33  & 2.1 & 6035$-$6870 \\
RSS & 2016 June 17  & 1800 & 0.26  & 2.00 & 120 & 2.1 & 6035$-$6870 \\
HRS & 2017 October 5  & 1500 & 0.04  & --   & --  & 3.0 & 3700$-$8900 \\
  \hline
\end{tabular}
\end{table*}

\begin{figure*}
\begin{center}
\includegraphics[width=12cm,angle=270,clip=]{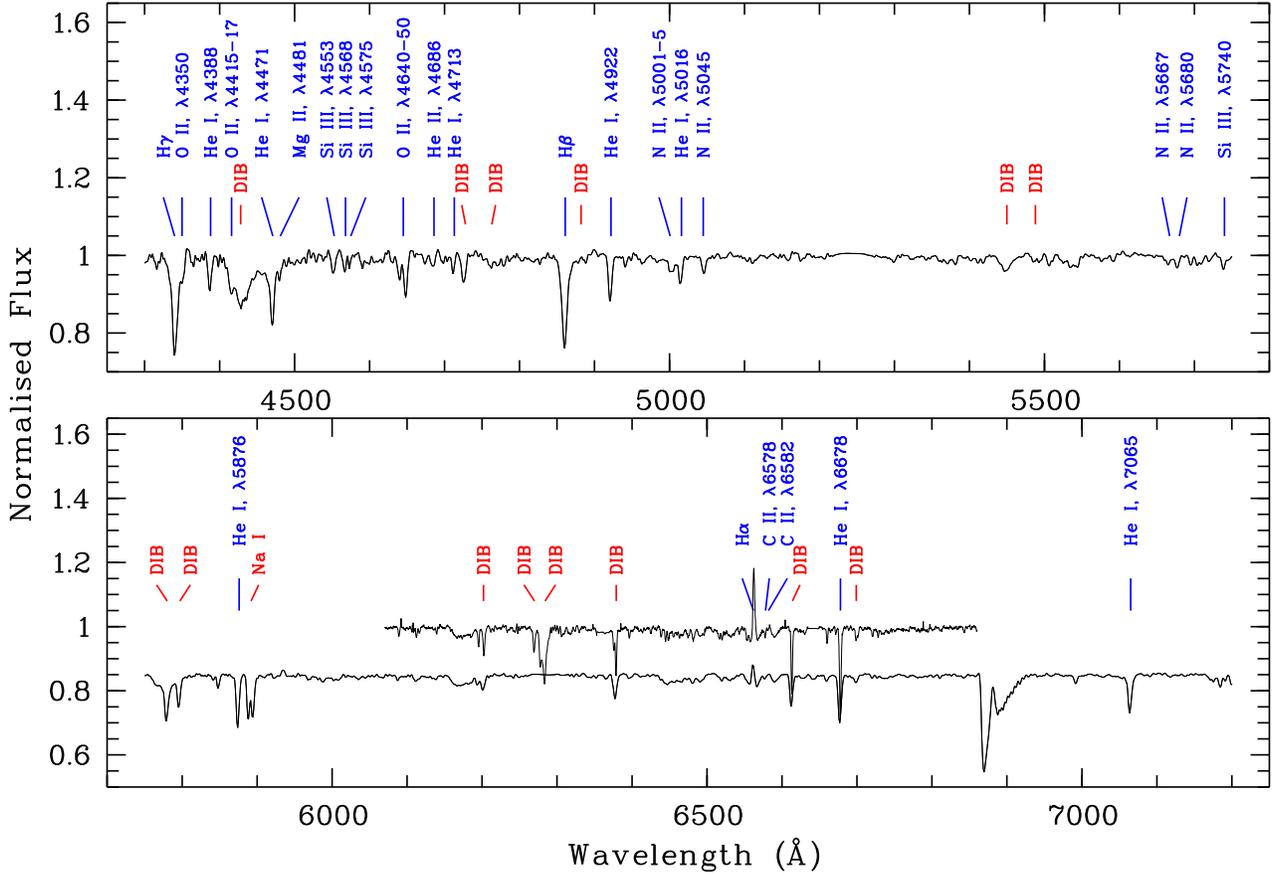}
\end{center}
\caption{Normalized low-resolution RSS spectrum of ALS\,19653 obtained with the SALT on 2016 April 25. The principal lines and 
most prominent DIBs are indicated. The lower panel also shows the high-resolution RSS spectrum (upper curve) obtained on 
2016 May 28.}
\label{fig:als}
\end{figure*}

Spectra of ALS\,19653 and the optical shells around it were taken with the Robert Stobie Spectrograph (RSS; Burgh et al. 2003; 
Kobulnicky et al. 2003) mounted on the Southern African Large Telescope (SALT; Buckley, Swart \& Meiring 2006; O'Donoghue et al. 
2006). Observations were carried out in the long-slit mode on three occasions during 2016 (logged in Table\,\ref{tab:obs}). 
The PG900 grating was used for the first observation (PA=0$\degr$) to cover the spectral range of 4200$-$7300~\AA\ with a final reciprocal 
dispersion of $0.97$~\AA \, pixel$^{-1}$. The spectral resolution FWHM was $4.43\pm$0.18~\AA. We call this spectrum low-resolution 
hereafter. An Xe lamp arc spectrum was taken immediately after the science frame. Spectrophotometric standard stars were observed 
during twilight time for the relative flux calibration. To study the radial velocity and H$\alpha$ intensity distributions in WS29 
and its surroundings, we obtained two additional spectra of higher resolution with the PG2300 grating and two different orientations 
of the slit. The slit was placed on ALS\,19653 and oriented in such a way to cross the brightest sides of the shell 
(PA=33$\degr$) and the protrusions in the
shell (PA=120$\degr$). The obtained spectra cover the spectral range of 6035$-$6870~\AA\ with a final reciprocal dispersion of 
$0.26$~\AA\ pixel$^{-1}$ and the spectral resolution FWHM of $1.85\pm$0.11~\AA. We call these spectra high-resolution hereafter.
A Ne lamp arc spectrum was taken immediately after science frames. The spatial scale of all three RSS spectra was 0.51 
arcsec pixel$^{-1}$.

Primary reduction of the RSS data was done in the standard way with the SALT science pipeline (Crawford et al. 2010). The subsequent
long-slit data reduction was carried out in the way described in Kniazev et al. (2008). The resulting reduced RSS spectrum 
of ALS\,19653 is shown in Fig.\,\ref{fig:als}, while those of WS29 are presented and discussed in Section\,\ref{sec:neb}.

To study ALS\,19653 in more detail and to search for possible radial velocity variations, we observed this star with the SALT High 
Resolution Spectrograph (HRS; Barnes et al. 2008; Bramall et al. 2010, 2012; Crause et al. 2014) on two occasions, 2016 May 27 and 
2017 October 5, with single exposures of 1500\,s and a seeing of about 3 arcsec (see Table\,\ref{tab:obs}). The HRS is a dual-beam, 
fibre-fed \'echelle spectrograph. It was used in the medium resolution mode (R=40\,000--43\,000 and 2.23 arcsec diameter for both the 
object and sky fibres) to obtain a spectrum in the blue and red arms over the total spectral range of $\approx$3700--8900~\AA. Both 
the blue and red arm CCDs were read out by a single amplifier with a 1$\times$1 binning. Three arc spectra of the ThAr lamp and three 
spectral flats were obtained for each observation in this mode during a weekly set of HRS calibrations.

Primary reduction of the HRS spectra was performed with the SALT science pipeline (Crawford et al. 2010). 
The subsequent reduction steps, including background subtraction, order extraction, removal of the blaze function, 
identification of the arc lines and merging of the orders for the object spectra, were carried out using the {\sc midas} HRS 
pipeline described in detail in Kniazev, Gvaramadze \& Berdnikov (2016). Parts of the 2016's HRS spectrum are shown in 
Fig.\,\ref{fig:als-mod}. 

\section{Spectral analysis and stellar parameters}
\label{sec:ana}

\subsection{Classification of BD+60$\degr$\,2668A\&B and ALS\,19653}
\label{sec:class}

\begin{figure*}
\includegraphics[width=14cm,angle=0]{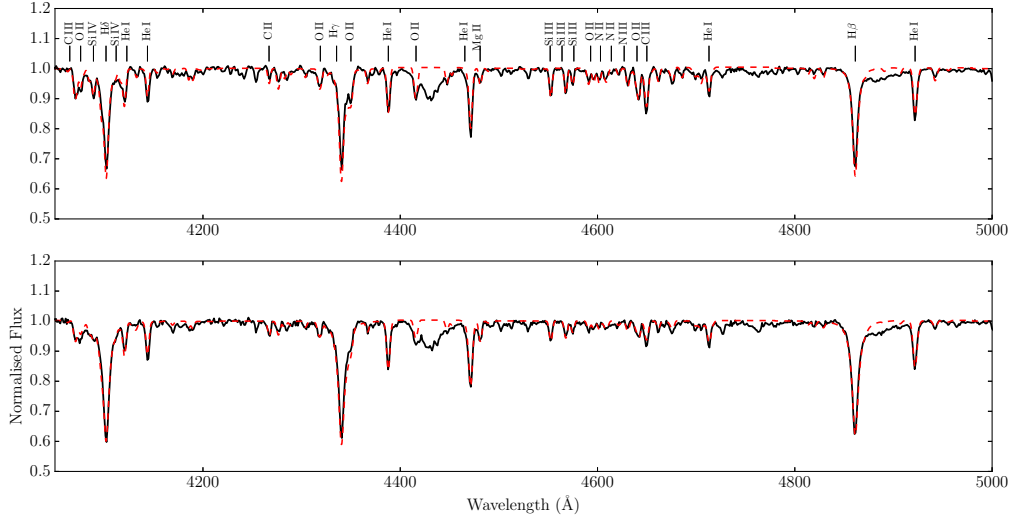}
\centering \caption{Parts of the normalized spectra of BD+60$\degr$\,2668A and BD+60$\degr$\,2668B (shown, respectively, in the 
upper and bottom panels by black lines) compared with the best-fitting {\sc fastwind} models (red dashed lines) with the parameters 
as given in Tables\,\ref{tab:par} and \ref{tab:abu}. The lines fitted by the models are indicated.} 
\label{fig:bd-mod}
\end{figure*}

\begin{figure*}
\includegraphics[width=14cm,angle=0]{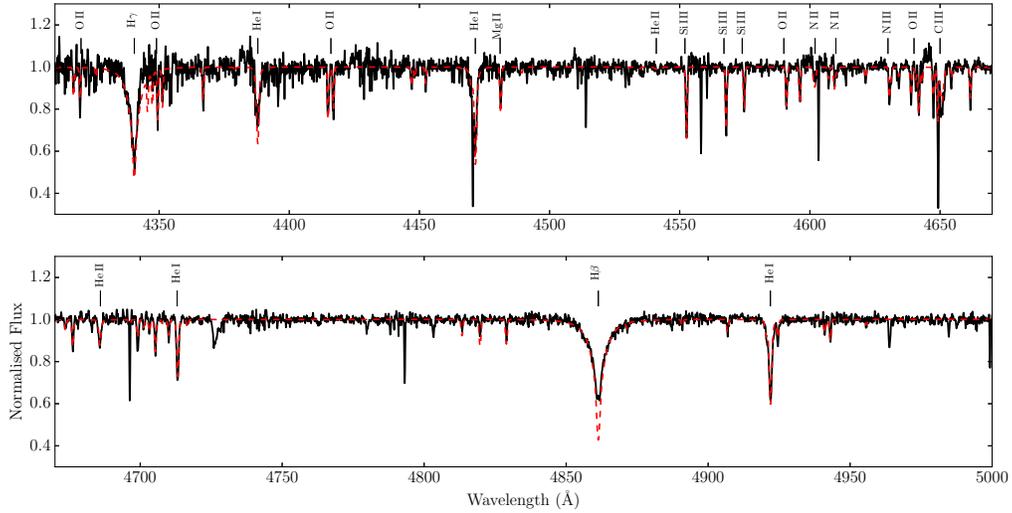}
\centering \caption{Parts of the normalized and re-binned SALT HRS spectrum of ALS\,19653 (black line) compared with the best-fitting 
{\sc fastwind} model (red dashed line) with the parameters as given in Tables\,\ref{tab:par} and \ref{tab:abu}. The lines fitted by 
the model are indicated.} 
\label{fig:als-mod}
\end{figure*}

The spectra of BD+60$\degr$\,2668A\&B and ALS\,19653 are dominated by the H\,{\sc i} and He\,{\sc i} absorption lines 
(see Figs\,\ref{fig:bd} and \ref{fig:als}). Two of these 
stars (BD+60$\degr$\,2668A and ALS\,19653) show the He\,{\sc ii} $\lambda$4686 line, implying that they are B stars of 
spectral types earlier than B0.7 (Walborn \& Fitzpatrick 1990). The HRS spectrum of ALS\,19653 also shows the presence of weak
He\,{\sc ii} $\lambda$4541 line. The only emission line detected is that of H$\alpha$ in the spectrum of ALS\,19653. 
We attribute the origin of this emission to the circumstellar material around the star (see next subsection). Note the presence
of a prominent diffuse interstellar band (DIB) at 6379\,\AA \, in the spectra of BD+60$\degr$\,2668A\&B, creating the illusion of 
asymmetry in the H$\beta$ line profile (cf. Figs\,\ref{fig:bd-mod} and \ref{fig:als-mod}).

Using the classification criteria from Evans et al. (2004), we classify BD+60\degr\,2668A and ALS\,19653 as 
B\,0.5\footnote{Although the He\,{\sc ii} $\lambda$4541 line was detected in the HRS spectrum of ALS\,19653, we do not take 
it into account because the classification criteria in Evans et al. (2004) are based on much lower resolution spectra.},
and BD+60\degr\,2668B as B\,1.5. The measured equivalent widths (EWs) of the H$\gamma$ line in the spectra of 
BD+60$\degr$\,2668A, BD+60$\degr$\,2668B and 
ALS\,19653 of $2.06\pm0.09$, $2.90\pm0.11$ and $1.42\pm0.04$ \AA, respectively, and the EW(H$\gamma$)-absolute magnitude 
calibration by Balona \& Crampton (1974) suggest for these stars the luminosity classes of Ib, III and Ia.
The luminosity classes derived for BD+60$\degr$\,2668A and ALS\,19653, however, are inconsistent with the absolute 
visual magnitudes of these stars of $M_V\approx-5.4$ mag and $-5.5$ mag (see Section\,\ref{sec:dis}), which rather
imply the luminosity classes of II and Ib, respectively (e.g. Humphreys \& McElroy 1984). We therefore regard BD+60$\degr$\,2668A 
and ALS\,19653 as B0.5\,II and B0.5\,Ib stars. 

\subsection{Spectral modelling}
\label{sec:mod}

Stellar atmosphere analysis of the three stars is rooted on a large grid of synthetic stellar atmosphere models built using the 
atmosphere/line formation code {\sc fastwind} (Santolaya-Rey, Puls \& Herrero 1997; Puls et al. 2005; Rivero Gonz{\'a}lez 
et al. 2012). The {\sc fastwind} stellar code takes into account non-local thermodynamic equilibrium effects in spherical symmetry 
with an explicit treatment of the stellar wind. The stellar grid was designed to accomplish the analysis of late O- and B-type stars 
with an effective temperature, $T_{\rm eff}$, between 34\,000 and 12\,000\,K in 1000\,K steps, and a surface gravity, $\log\,g$, 
between 4.4 and 2.0\,dex in steps of 0.1\,dex. The helium abundance was fixed at the solar value. Explicit atomic models for  
H\,{\sc i}, He\,{\sc i,ii},  N\,{\sc ii,iii}, O\,{\sc ii,iii}, C\,{\sc ii,iii}, Si\,{\sc ii,iii,iv} and Mg\,{\sc ii} were included 
in the determination of the fundamental stellar parameters and chemical abundances. The rest of the chemical species 
were treated in an implicit way to account for blanketing/blocking effects. For further details see Puls et al. (2005).

First, several key lines in the spectra\footnote{For spectral modelling of ALS\,19653, we used the HRS spectrum obtained on 2016 May 
25.} were simultaneously compared with the grid looking for the set of stellar parameters that best reproduce the spectrum, following 
the routines and the line list described in Castro et al. (2012; see also Lefever et al. 2010). Subsequently, and based on the best 
predicted effective temperatures and gravities, new synthetic sub-grids were built for each star with chemical abundances varied 
around the cosmic abundance standard (CAS) in the solar neighbourhood (Nieva \& Przybilla 2012) in steps of 0.2\,dex, and 
microturbulence velocities, $\xi$, spanning from 1 to 15~\kms \, in steps of 1~\kms. The best combination of abundances and
microturbulence that reproduce the observations were found through an optimized genetic algorithm. The main chemical transitions 
modelled in this study are labelled in Figs\,\ref{fig:bd-mod} and \ref{fig:als-mod} (see also table\,4 in Castro et al. 2012). The 
derived stellar parameters and chemical abundances are listed in Tables\,\ref{tab:par} and \ref{tab:abu}.

\begin{table}
\caption{Stellar parameters for BD+60$\degr$\,2668A\&B  and ALS\,19653.}
\label{tab:par}
\begin{center}
\begin{tabular}{lccc}
\hline
                    & BD+60$\degr$\,2668A     & BD+60$\degr$\,2668B     & ALS\,19653   \\
\hline
$T_{\rm eff}$ (kK)  & $26.0^{+1.7} _{-1.8}$   & $23.0^{+1.4} _{-1.2}$   & $26.0^{+1.3} _{-1.7}$ \\
$\log g$            & $3.30^{+0.14} _{-0.15}$ & $3.60^{+0.17} _{-0.12}$ & $3.10^{+0.26} _{-0.28}$ \\
$\xi (\kms)$        & $9\pm2$                 & $8\pm2$                 & $7\pm2$  \\
$v\sin i$ ($\kms)$    & 130                     & 150                     & 28           \\
$v_{\rm r,hel}$ ($\kms$)  & $-20.9\pm2.1$     & $-20.9\pm3.6$           & see Table\,\ref{tab:RV} \\
\hline
\end{tabular}
\end{center}
\end{table}

\begin{table*}
\caption{Elemental abundances (by number) in BD+60$\degr$\,2668A\&B and ALS\,19653. The cosmic abundance standard (CAS; Nieva \& 
Przybilla 2012) in the solar neighbourhood and initial abundances adopted in the evolutionary models by Brott et al. (2011) are given 
for reference. Note the enhanced N abundance in BD+60$\degr$\,2668A.}
\label{tab:abu}
\begin{center}
\begin{tabular}{lccccc}
\hline
	            & BD+60$\degr$\,2668A  & BD+60$\degr$\,2668B  & ALS\,19653          & CAS           & Brott et al. (2011) \\ 
\hline
$\log$(C/H)+12  & 7.7$^{+0.2}_{-0.2}$  & 7.5$^{+0.2}_{-0.2}$  & 7.9$^{+0.5}_{-0.2}$ & $8.33\pm0.04$ & 8.13 \\  
$\log$(N/H)+12  & 8.3$^{+0.2}_{-0.2}$  & 7.9$^{+0.1}_{-0.2}$  & 7.9$^{+0.2}_{-0.2}$ & $7.79\pm0.04$ & 7.64 \\  
$\log$(O/H)+12  & 8.7$^{+0.1}_{-0.1}$  & 8.8$^{+0.1}_{-0.1}$  & 8.5$^{+0.2}_{-0.2}$ & $8.76\pm0.05$ & 8.55 \\  
$\log$(Mg/H)+12 & 7.6$^{+0.4}_{-0.5}$  & 7.7$^{+0.4}_{-0.2}$  & 7.5$^{+0.3}_{-0.1}$ & $7.50\pm0.05$ & 7.32 \\ 
$\log$(Si/H)+12 & 7.6$^{+0.2}_{-0.3}$  & 7.5$^{+0.2}_{-0.1}$  & 7.5$^{+0.2}_{-0.1}$ & $7.56\pm0.05$ & 7.41 \\
\hline
\end{tabular}
\end{center}
\end{table*}

The high resolution of the HRS spectrum of ALS\,19653 allowed us to determine the projected rotational velocity, $v\sin i$, 
through the Fourier transform of the Si\,{\sc iii}\,$\lambda$4552 line profile using the {\sc iacob-broad} code (Sim\'on-Diaz \& 
Herrero 2007, 2014). The much lower spectral resolution of the TWIN spectra prevents us from applying the same 
technique for BD+60$\degr$\,2668A\&B. For these stars, after a first guess of the best stellar parameters, we convolved the 
synthetic spectra until we reproduced the FWHM of the metallic lines. This procedure was repeated, updating $v\sin i$,  
until it converged and we reached the best combination of stellar parameters and $v\sin i$ that fits the observations. 
The obtained values of $v \sin i$ are given in Table\,\ref{tab:par}. We note that the actual values of $v \sin i$ derived for 
BD+60$\degr$\,2668A\&B could be smaller because the measured velocities are around the limit imposed by the low spectral resolution.

We also estimated the heliocentric radial velocities, $v_{\rm r,hel}$, of all three stars using the {\sc ulyss} (University 
of Lyon Spectroscopic analysis Software) program (Koleva et al. 2009) with a medium spectral-resolution library (Prugniel, 
Vauglin \& Koleva 2011). For BD+60$\degr$\,2668A\&B we derived almost equal velocities (see Table\,\ref{tab:par}).
For ALS\,19653 we used all five RSS and HRS spectra and found significant radial velocity 
variations ($\Delta v_{\rm r,hel}\approx50$~\kms; see Table\,\ref{tab:RV}). These changes suggest that ALS\,19653 is a close
binary system, which was in the periastron passage between 2016 April 25 and May 27. 

\begin{table}
  \caption{Heliocentric radial velocity changes with time in the spectrum of ALS\,19653.}
  \label{tab:RV}
  \begin{center}
   \begin{tabular}{lcc}
      \hline
Date & $v_{\rm r,hel} \, (\kms)$ & Spectrograph \\
\hline
2016 April 25  & $-29\pm3$         & RSS GR900  \\
2016 May 27    & $+9.25\pm0.33$    & HRS        \\
2016 May 28    & $+10.2\pm0.5$     & RSS GR2300 \\
2016 June 17   & $+23.7\pm1.3$     & RSS GR2300 \\
2017 October 5 & $+12.18\pm1.33$   & HRS        \\
\hline
    \end{tabular}
    \end{center}
    \end{table}

\begin{figure}
\begin{center}
\includegraphics[width=6cm,angle=270,clip=]{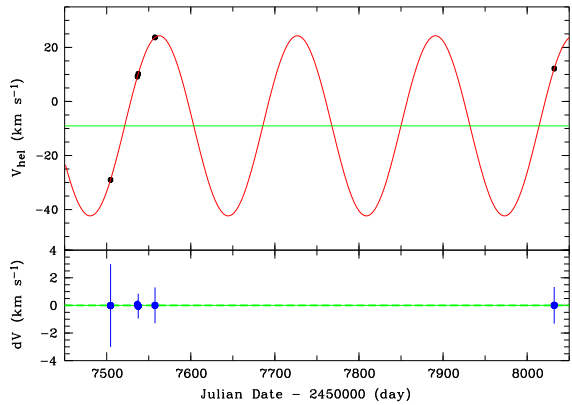}
\end{center}
\caption{Radial velocity changes with time in the spectrum of ALS\,19653.}
\label{fig:RV}
\end{figure}

To get an idea on possible parameters of 
the binary system, we assume that its orbit is circular and fit the data points to a sine curve using the $\chi ^2$ algorithm. 
The best-fitting result ($\chi ^2=2.33\times10^{-3}$) is shown in Fig.\,\ref{fig:RV} and implies an orbital period of 
$164.20\pm0.02$\,d, an amplitude of $33.33\pm0.04$~\kms \, and a systemic velocity of $-9.02\pm0.05$~\kms. We realize that
this solution is not unique because of the limited number of data points. Time-series spectroscopy and photometry of ALS\,19653 
are required to determine the actual parameters of the system.

Still, using the above parameters of the binary and the mass of ALS\,19653 of $\approx20 \, \msun$ (see 
Section\,\ref{sec:ws29}), one can estimate the mass of the companion star to be $\approx8 \, \msun$. Assuming that this star is
on the main sequence, one finds that its $V$-band brightness should be a factor of $\sim20-30$ lower than that of ALS\,19653. This 
is consistent with the non-detection of double-lined structures in the HRS spectra of ALS\,19653, although both spectra were 
obtained at the time favourable for their detection, i.e. when the difference of radial velocities of the binary 
components was close to maximum (cf. Fig.\,\ref{fig:RV} and Table\,\ref{tab:RV}). 

\begin{figure}
\begin{center}
\includegraphics[angle=0,width=0.5\textwidth,clip=]{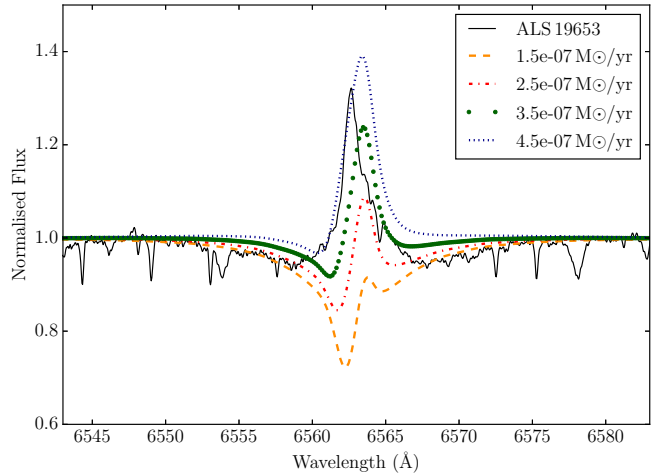}
\end{center}	
\caption{Comparison of the observed profile of the H$\alpha$ line with synthetic profiles predicted	by {\sc fastwind} models 
with different values of the mass-loss rate as specified in the legend.}
\label{fig:wind}
\end{figure}

\begin{figure}
\includegraphics[width=7cm,angle=270]{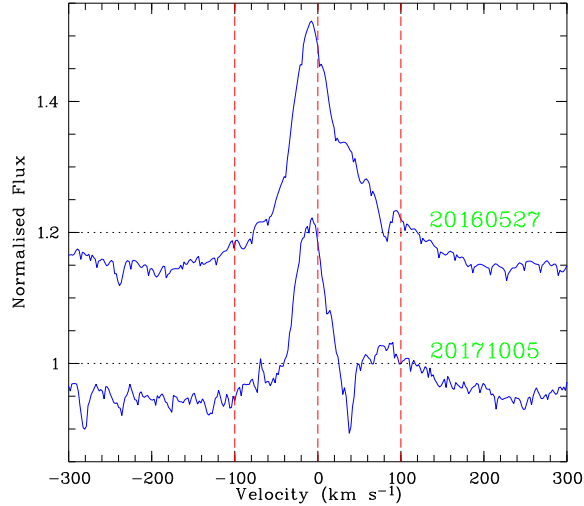}
\centering \caption{Changes in the H$\alpha$ line profile in the HRS spectra of ALS\,19653. 
The upper line profile corresponds to the spectrum from 2016 May 27, while the lower one to the spectrum 
from 2017 October 5.}
\label{fig:prof}
\end{figure}

The presence of the H$\alpha$ emission line in the spectrum of ALS\,19653 (see Fig.\,\ref{fig:als}) could indicate that this 
star possesses a strong wind. In Fig.~\ref{fig:wind}, we compare the observed H$\alpha$ line profile (from the HRS spectrum
obtained on 2016 May 27) with synthetic ones obtained in {\sc fastwind} models with four different mass-loss rates. In these models, 
we adopted a wind velocity law exponent $\beta$=2 and a terminal wind velocity of 1500~\kms, based on the effective temperature of
ALS\,19653 and the empirical calibration in Kudritzki \& Puls (2000). The intensity of the observed line matches a mass-loss rate 
of $\approx4\times10^{-7} \, \myr$. However, the synthetic lines have a P\,Cygni type profile (typical of lines formed in powerful 
winds or expanding gaseous shells), while in the observed line the intensity peak is shifted to the violet side with respect to the 
peaks of the model lines and the line itself has an asymmetric shape with an absorption dip on the red side (see also
Fig.\,\ref{fig:prof}). This suggests that the H$\alpha$ emission is rather formed in the circumstellar material around the star.

The circumstellar origin of the H$\alpha$ line is also suggested by changes in the profile of this line.
Fig.\,\ref{fig:prof} shows that in 2016 the H$\alpha$ line was dominated by a sharp asymmetric emission
and shows also the presence of a second, much 
weaker emission component at $\approx90$ \, \kms, separated from the first one by a shallow absorption dip at $\approx80$ \kms. 
In 2017 the two-peak shape of the H$\alpha$ line became more obvious after the absorption dip has extended below the continuum and 
shifted bluewards to $\approx40$ \kms. This change in the absorption was accompanied by $\approx10$ per cent decrease of the 
intensity of the main emission component and a blueward shift of the second emission peak.
Similar behaviour of the H$\alpha$ emission line was detected in late B and early A supergiants (e.g. Kaufer et al. 1996; Verdugo, 
Talavera \& G\'{o}mez de Castro 2000; Markova \& Valchev 2000), where the shape of this line changes from blue- or red-shifted 
asymmetric emission to double-peaked or inverse P\,Cygni profile (see, e.g., fig.\,1 in Markova \& Valchev 
2000 for a good example of these changes). This behaviour of the line profile was interpreted in terms of deviation of stellar 
wind from spherical symmetry or presence of a flattened (disc-like) circumstellar structure (e.g. Kaufer et al. 1996; Petrenz 
\& Puls 1996; Fullerton et al. 1997). The presence of elongated (IR) nebula around ALS\,19653 and the detection of 
intrinsic polarization of this star (with the polarization angle aligned with the elongated nebula) makes it natural to assume that 
the H$\alpha$ line originates in a flattened circumstellar structure (viewed nearly edge-on) and that its variability is caused 
by a large-scale density asymmetry in this structure (cf. Section\,\ref{sec:ws29}).

\begin{figure}
\begin{center}
\includegraphics[angle=0,width=0.2\textwidth,clip=]{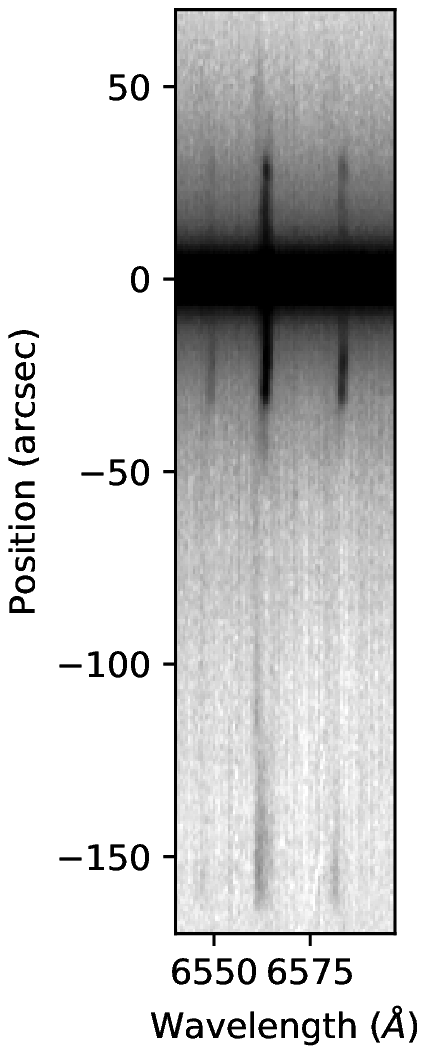}
\includegraphics[angle=0,width=0.253\textwidth,clip=]{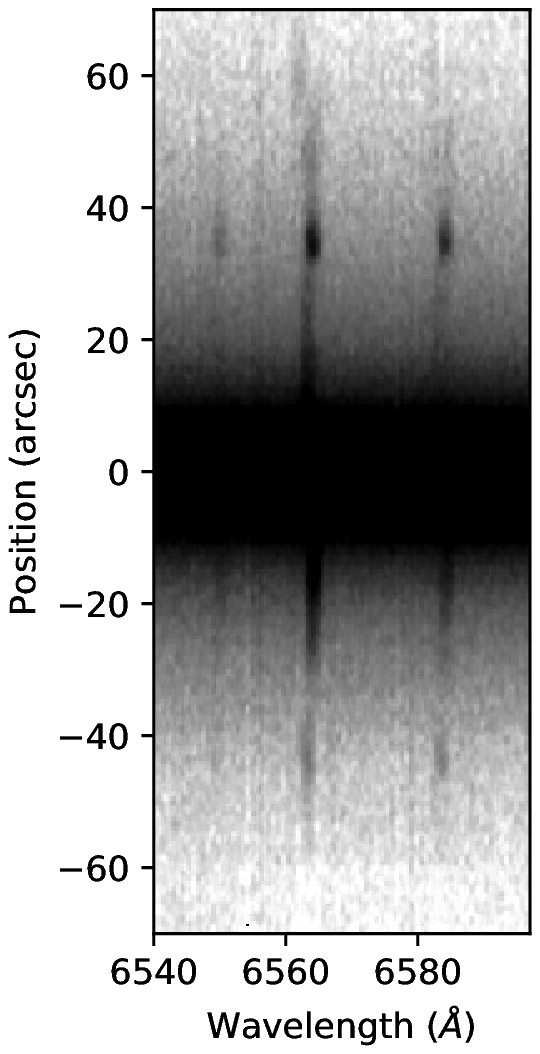}
\end{center}
\caption{Portions of the 2D high-resolution RSS spectra of WS29, showing the H$\alpha$ and [N\,{\sc ii}] $\lambda\lambda$6548, 6584
lines. The left- and right-hand panels correspond, respectively, to the slit orientation of PA=33$\degr$ and 120$\degr$. Negative 
offsets in the left-hand panel correspond to the north-east side of the slit, while in the right-hand panel to the south-east side
of the slit.} 
\label{fig:2D}
\end{figure}

\begin{figure*}
\begin{center}
\includegraphics[width=7cm,angle=0,clip=]{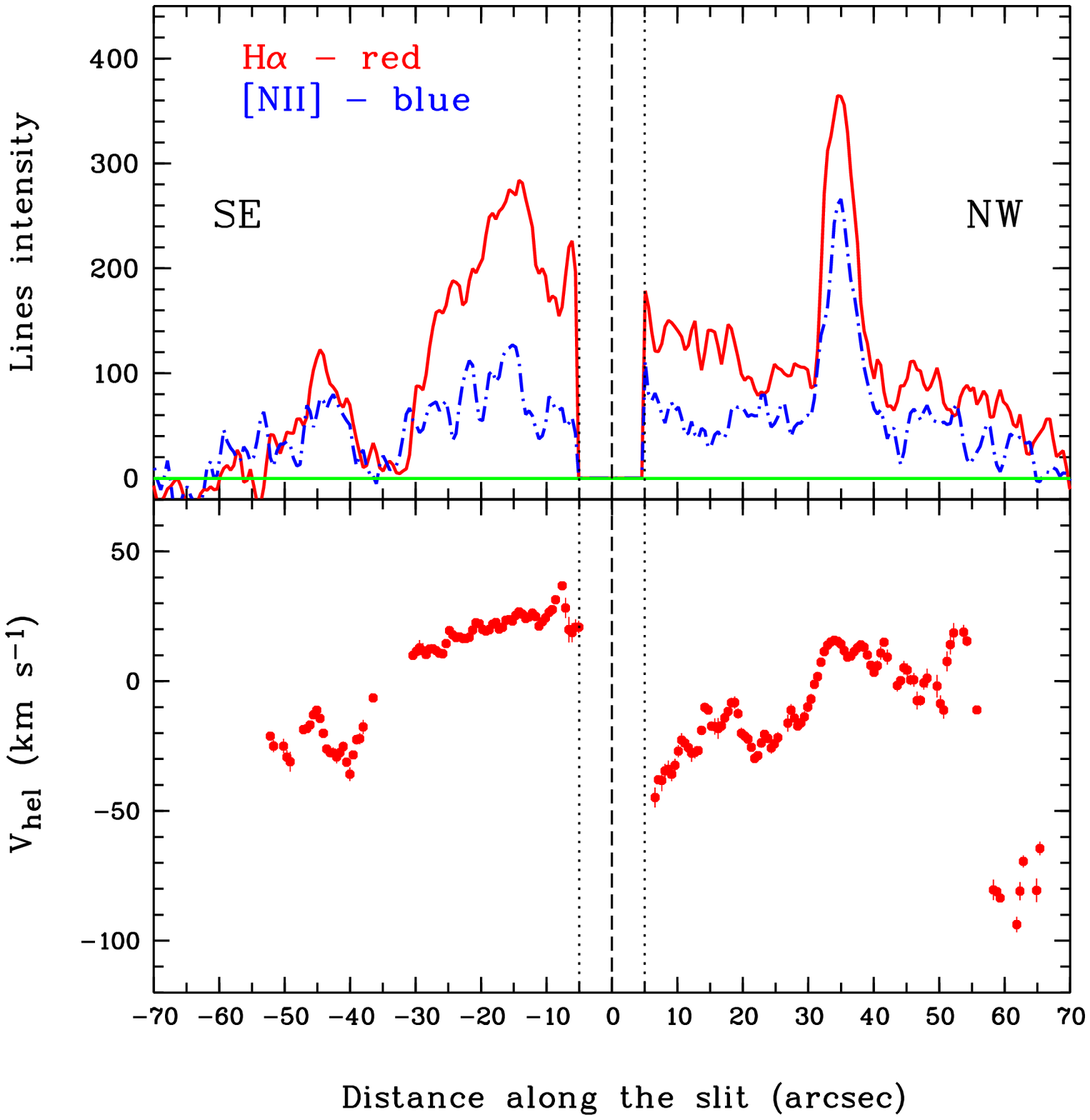}
\includegraphics[width=7cm,angle=0,clip=]{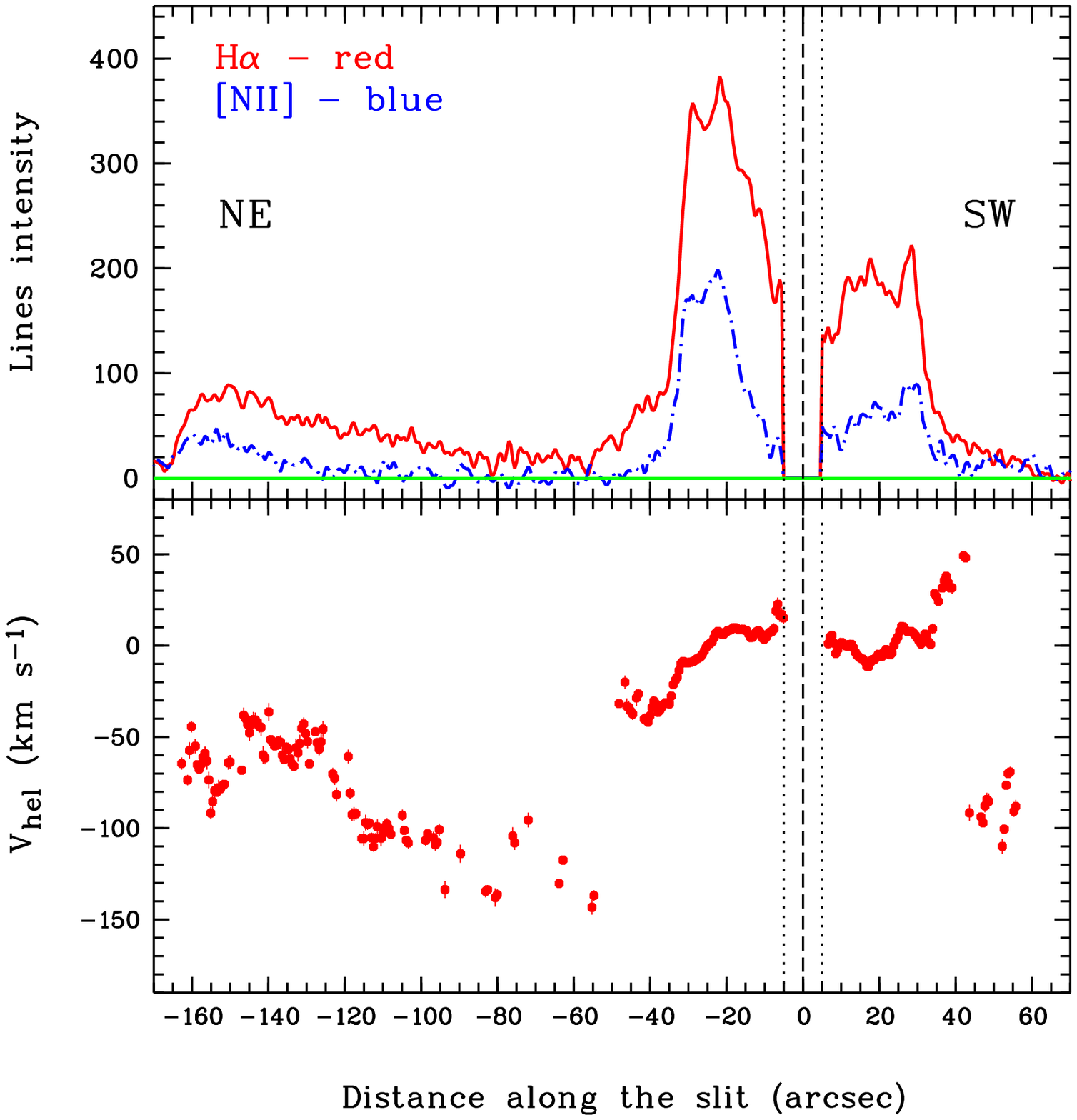}
\includegraphics[width=14.5cm,angle=0,clip=]{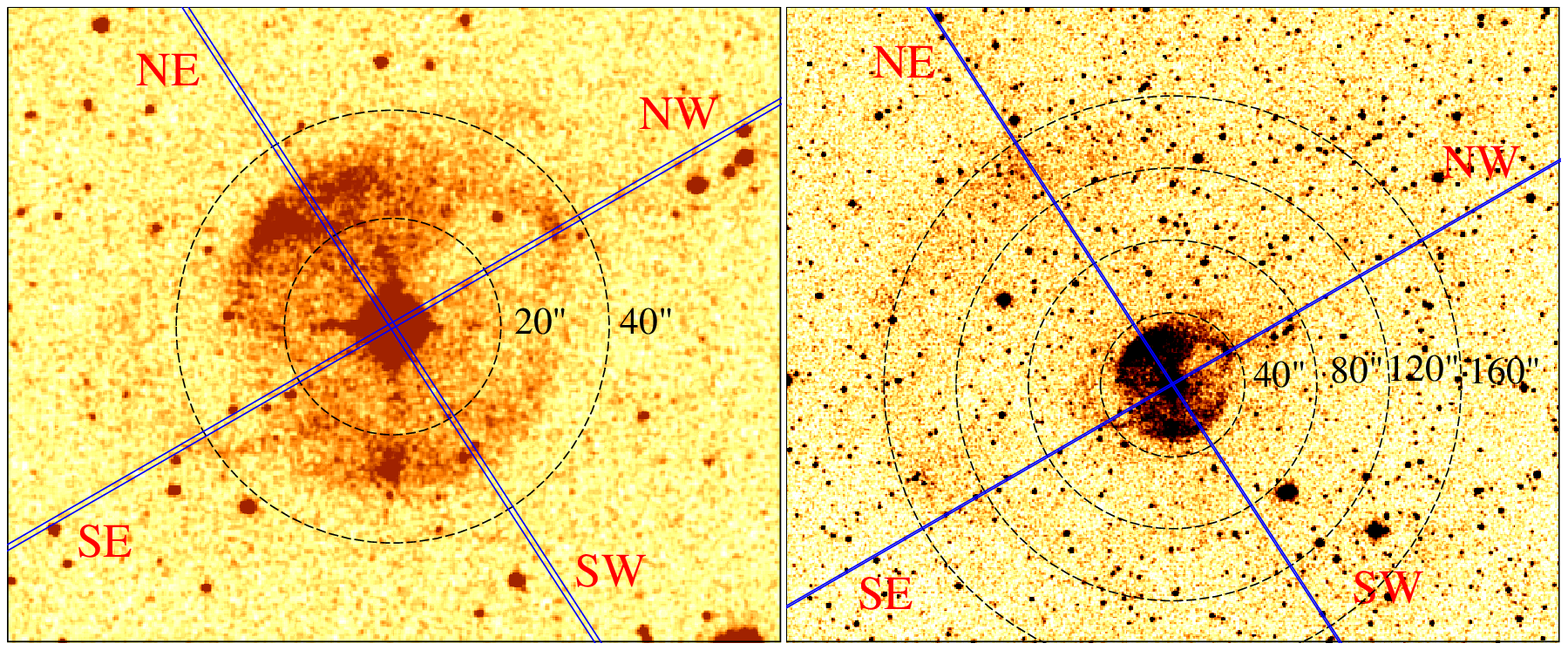}
\end{center}
\caption{Upper panels: Distributions of the H$\alpha$ and [N\,{\sc ii}] $\lambda$6584 emission line intensities and H$\alpha$ 
heliocentric radial velocity, $V_{\rm hel}$, in the high-resolution RSS spectra of WS29 and its surroundings along the slits with 
PA=120$\degr$ (left-hand panel) and 33$\degr$ (right-hand panel). The dashed vertical line in both panels corresponds to the 
position of ALS\,19653, while the dotted vertical lines at $\pm5$ arcsec from the central line mark the area where the line 
intensities and velocity were not measured because of the effect of the star. The directions of the slits are shown. Bottom 
panels: IPHAS H$\alpha$ images of WS29 and its surroundings with the slit positions shown by blue rectangles of angular width of 
2 arcsec. Concentric, dashed circles are overplotted on the images to facilitate their comparison with the upper panels.}
\label{fig:ha}
\end{figure*}

\begin{figure}
\begin{center}
\includegraphics[angle=-90,width=12cm,clip=]{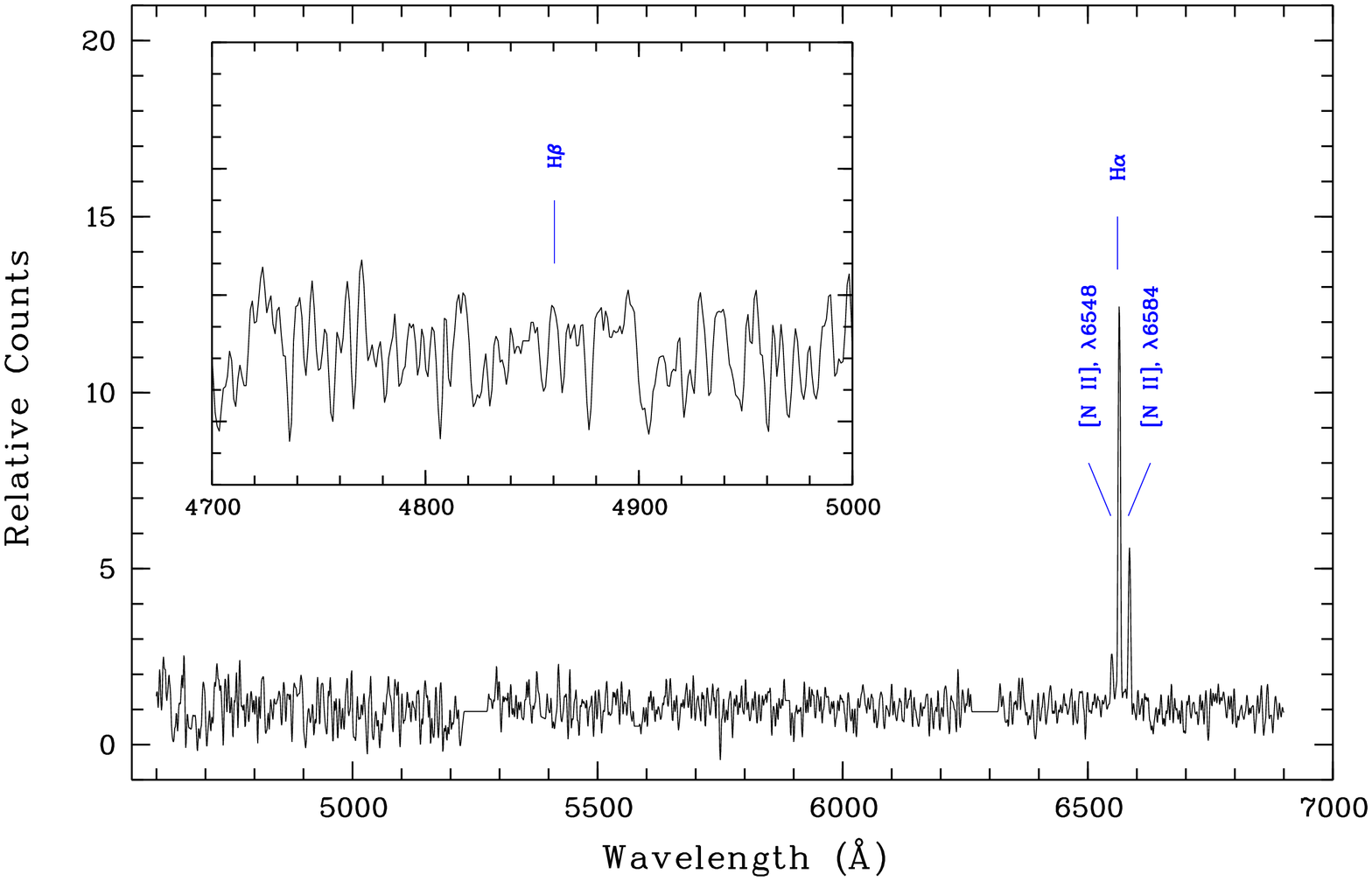}
\end{center}
\caption{1D low-resolution RSS spectrum of WS29. The insert shows a part of the spectrum around the
H$\beta$ line.}
\label{fig:1D}
\end{figure}

\section{Spectroscopy of WS29}
\label{sec:neb}

As noted in Section\,\ref{sec:obs}, we obtained two high-resolution RSS spectra of WS29 with the slit placed on ALS\,19653 and 
oriented at PA=33$\degr$ and 120$\degr$. The 8 arcmin long slit allowed us also to obtain spectra of the second, more extended shell 
around ALS\,19653 (see Fig.\,\ref{fig:ws29-2}). The only lines detected in the spectra of WS29 and the second shell are those of 
H$\alpha$ and [N\,{\sc ii}] $\lambda\lambda$6548, 6584. Portions of the 2D spectra showing these lines are presented in 
Fig.\,\ref{fig:2D}. The left-hand panel shows that the second shell extends in the north-east direction for $\approx160$ arcsec 
from the star and that the emission lines from the shell are split in two components, indicating expansion of the shell.
The splitting of the emission lines is also visible in the right-hand panel (at about 60 arcsec southwest of the star),
indicating the presence of material with different radial velocities (cf. Fig.\,\ref{fig:ha}).

Using the 2D spectra, we plot the distributions of the H$\alpha$ and [N\,{\sc ii}] $\lambda$6584 emission line 
intensities and the H$\alpha$ heliocentric radial velocity, $V_{\rm hel}$, along the slits (see the upper panels in Fig.\,\ref{fig:ha}, 
where the left- and right-hand panels correspond, respectively, to the slit orientation of PA=120$\degr$ and 33$\degr$). The 
plots show that the intensities of both lines correlate with bright details in WS29 and the second shell. In particular, the plots also 
show that WS29 is more extended in the southeast-northwest direction (PA=120$\degr$) than in the orthogonal direction.

The position-velocity diagram obtained for the slit with PA=33$\degr$ shows that the nearby side of the second shell approaches us with a
velocity of $\sim100$~\kms. The velocity field in WS29 is more complex and shows different behaviour along the two slits. The velocity 
distribution along the slit with PA=33$\degr$ (i.e. along the long axis of the IR nebula) shows that the northeast
rim of the optical shell is approaching us with a velocity of about $20-30$~\kms, while the rim of the opposite side of the
shell is receding from us with about the same velocity. This suggests that the outer parts of WS29 might be rotating around an 
axis aligned with the slit at PA=120$\degr$ (we discuss this possibility in Section\,\ref{sec:ws29}). The velocity distribution along the 
slit with PA=120$\degr$ indicates that we see an expanding bilobal structure, whose existence also follows from the presence of filaments
protruding in the southeast and northwest directions beyond the main (almost circular) optical shell (the origin of these protrusions is 
discussed in Section\,\ref{sec:ws29} as well).

Fig.\,\ref{fig:1D} shows the 1D low-resolution RSS spectrum of WS29 obtained by summing up, without any weighting, all rows from 
the area of an annulus with an outer radius of 40 arcsec centred on ALS\,19653 and the central $\pm5$ arcsec excluded. The emission 
lines (H$\beta$, H$\alpha$ and [N\,{\sc ii}] $\lambda\lambda$6548, 6584) detected in the resulting spectrum were measured using the 
programs described in Kniazev et al. (2004). Table\,\ref{tab:int} lists the observed intensities of these lines normalized to 
H$\beta$, $F(\lambda)/F$(H$\beta$), the reddening-corrected line intensity ratios, $I(\lambda)/I$(H$\beta$), and the logarithmic 
extinction coefficient, $C$(H$\beta$), which corresponds to $E(B-V)=1.94\pm0.56$ mag. This value of $E(B-V)$ agrees within
the error margins with the colour excess of ALS\,19653 of $E(B-V)=2.22\pm0.06$ mag based on the $B$ and $V$ magnitudes of this star
(see Table\,\ref{tab:det}) and the intrinsic colour of a B0.5\,Ib star of $(B-V)_0=-0.22$ mag (Fitzgerald 1970).
\begin{table}
\centering{
\caption{Line intensities in the low-resolution RSS spectrum of WS29.}
\label{tab:int}
\begin{tabular}{lcc} \hline
\rule{0pt}{10pt}
$\lambda_{0}$(\AA) Ion & F($\lambda$)/F(H$\beta$)&I($\lambda$)/I(H$\beta$) \\ \hline
4861\ H$\beta$\        &  $1.00\pm0.61$  & $1.00\pm0.64$ \\
6548\ [N\ {\sc ii}]\   &  $3.23\pm2.33$  & $0.37\pm0.30$ \\
6563\ H$\alpha$\       & $26.17\pm18.92$ & $2.94\pm2.13$ \\
6584\ [N\ {\sc ii}]\   & $10.33\pm6.80$  & $1.13\pm0.85$ \\
  & & \\
$C$(H$\beta$)          & \MC {2}{c}{$2.86\pm0.82$} \\  
$E(B-V)$               & \MC {2}{c}{$1.94\pm0.56$ mag} \\
\hline
\end{tabular}
 }
\end{table}

\section{Discussion}
\label{sec:dis}

\subsection{WS3 and BD+60$\degr$\,2668A\&B}
\label{sec:ws3}

The similar distances to and radial velocities of BD+60$\degr$\,2668A\&B, and the small angular (or projected linear) 
separation between these stars of $\approx3$ arcsec (or $\approx0.05$ pc) strongly suggest that they are members of the
same stellar system. We note that both stars are located within the boundaries of the Cas\,OB5 association (Humphreys 1978) and 
that the {\it Gaia} DR2 parallaxes of a large part of stars listed in Humphreys (1978; see there table\,8) as the association members 
are similar to those of BD+60$\degr$\,2668A\&B. This implies that BD+60$\degr$\,2668A\&B reside in Cas\,OB5 and justifies why 
we adopt the same distance to both stars (see Section\,\ref{sec:two}).

Using the adopted distance of 3.6\,kpc, the $B$ and $V$ photometry from Table\,\ref{tab:det}, $T_{\rm eff}$ from 
Table\,\ref{tab:par}, $(B-V)_0$ colours and bolometric corrections from Fitzgerald (1970), and assuming that the 
total-to-selective absorption ratio equals to 3.1, we calculate absolute visual and bolometric magnitudes: $M_V\approx-5.4$ mag 
and $M_{\rm bol}\approx-7.9$ mag for BD+60$\degr$\,2668A and $M_V\approx-4.6$ mag and $M_{\rm bol}\approx-6.8$ mag for 
BD+60$\degr$\,2668B. The derived values of $M_{\rm bol}$ translate into the bolometric luminosities of BD+60$\degr$\,2668A\&B of 
$\log(L/\lsun)\approx5.0$ and 4.6, which along with the effective temperatures of these stars 
imply that both of them are still on the main sequence or only recently left it, and that their initial (zero-age 
main-sequence) masses were, respectively, $M_{\rm ZAMS}\approx20$ and $15 \, \msun$ (e.g. Brott et al. 2011).

The inference that both BD+60$\degr$\,2668A\&B are relatively unevolved is supported by the surface elemental abundances 
measured for these stars. The comparison of the CNO abundances of BD+60$\degr$\,2668A given in Table\,\ref{tab:abu} with those predicted 
for $20 \, \msun$ rotating main-sequence stars by the stellar evolutionary models of Brott et al. (2011) shows that they agree 
with each other well enough if the initial rotational velocity of BD+60$\degr$\,2668A was $\sim400$~\kms, meaning that the
enhanced nitrogen abundance of this star could be caused by rotational mixing. The only difference is that the Brott et al.'s models
predict a 0.2 dex lower O abundance, which could be due to the lower initial abundance for this element (as compared to the CAS) adopted 
in these models (cf. Table\,\ref{tab:abu}). The Brott et al.'s models also suggest that the age of BD+60$\degr$\,2668A is 
$\approx7-8$ Myr. Similarly, we found that the CNO abundances of BD+60$\degr$\,2668B would reasonably agree with the model predictions if 
the initial rotational velocity of this star was $\sim200$~\kms. The models also suggest that the age of BD+60$\degr$\,2668B is
$\approx11$ Myr, i.e. a factor of $\approx1.5$ higher than that of BD+60$\degr$\,2668A. 

The age estimates for BD+60$\degr$\,2668A\&B are in seeming contradiction with the possibility that the two stars are
members of the same stellar system. The age discrepancy, however, could be understood if BD+60$\degr$\,2668A is a blue straggler,
i.e. a rejuvenated product of binary mass transfer or merger, implying that this star is or was a binary system.
This possibility appears reasonable because the majority of massive stars are formed in binary and multiple systems and because 
their evolution is dominated by various binary interaction processes (Sana et al. 
2012). Moreover, the binary population synthesis modelling shows (de Mink et al. 2014) that about 30 per cent 
of massive stars undergo binary interaction (i.e. mass transfer, common envelope evolution and/or merger) during the main sequence 
stage. A factor of several enhancement of the nitrogen abundance on the surface of BD+60$\degr$\,2668A could be the result of 
binary interaction as well (cf. Langer 2012).

Figure\,\ref{fig:ws3} shows that the circumstellar nebula around BD+60$\degr$\,2668A\&B
is composed of two distinct components: a diffuse halo and an inner shell. Similar two-component nebulae were also detected around two 
candidate LBVs, GAL\,079.29+00.46 (Gvaramadze et al. 2010b; see their fig.\,2j) and [GKF2010]\,MN112 (Gvaramadze et al. 2010a; 
see their fig.\,1). The morphology of these nebulae might be interpreted as indicating that their central stars experienced a brief 
episode of enhanced mass loss, leading to the formation of a compact region of dense material (visible in H$\alpha$ as a density-bounded 
\hii region --- the halo), and that afterwards the stellar wind increases its speed and sweeps up the material of the preceding slower 
wind, and thereby creates a shell within the halo.

In the life of a single massive star such changes in the mass-loss rate and wind velocity can occur if after the red supergiant stage 
the star becomes a Wolf-Rayet star or undergoes a blue loop evolution. The CNO abundances derived for BD+60$\degr$\,2668A\&B, however, 
suggest that these stars have not yet gone through the red supergiant stage. Besides this, the episode of enhanced mass loss and 
subsequent increase of the wind velocity could be caused by the bi-stability jump (Pauldrach \& Puls 1990; Lamers \& Pauldrach 1991)
if in the course of stellar evolution the effective temperature falls below some critical value ($\approx21-22$\,kK) and then again 
raises above it. According to the stellar evolutionary models by Brott et al. (2011), for $15-20 \, \msun$ stars this situation takes 
place during the transition from core-H to core-He burning. The duration of this transition of $\sim10^5$\,yr, however, is too long and 
the wind velocity is too high ($\sim1000$~\kms) to result in the origin of a compact (parsec-size) circumstellar nebula.

On the other hand, as discussed above, 
BD+60$\degr$\,2668A might be the result of merger of two stars, which causes a temporal inflation of the resulting 
single star and a decrease of its $T_{\rm eff}$ below the temperature of the bi-stability jump. As a result of this, the stellar 
wind velocity drops by a factor of 2, while the mass-loss rate increases by a factor of 10 (Vink 2018), leading to a 
factor of 20 increase of the stellar wind density and formation of a halo around the star. Later on, after the thermal 
adjustment of the merger product on its Kelvin-Helmholtz time-scale ($\sim10^4$ yr), the effective temperature of the star increases 
above the temperature of the bi-stability jump and the stellar wind velocity increases to its initial value, which in turn leads to 
the formation of a shell within the halo. Since the (relaxed) merger product is expected to be a fast-rotating star, the temperature 
distribution across its surface could be inhomogeneous due to gravity darkening. Correspondingly, the wind velocity along the 
stellar rotational axis might be higher than in the equatorial region, which could be responsible for the observed elongated 
shape of the shell (see Fig.\,\ref{fig:ws3}). We speculate also that the post-merger relaxation of the newly formed single star might 
be accompanied by photometric and spectroscopic variability typical of LBVs and that in the recent past BD+60$\degr$\,2668A probably 
looked like an LBV.

The possibility that the LBV activity (including giant, $\eta$\,Car-like, eruptions) is triggered by the interaction between companion 
stars in binary or triple systems (e.g. through the merger of two stars, tidal interaction, or interchange of components) has been 
widely discussed in the last two decades (e.g. Justham, Podsiadlowski \& Vink 2014; Portegies Zwart \& van den Heuvel 2016, and 
references therein). Since this interaction can take place at both early and late 
stages of evolution of binary/triple systems, it is natural to expect that the LBV phenomenon (accompanied by the formation of 
circumstellar nebulae) can be found among massive stars at both early and advanced evolutionary stages. It should also be noted that the 
detection of possible companion stars in wide orbits around several (candidate) LBVs (e.g. Martayan et al. 2016) does not contradict 
the possibility that the LBV activity of these stars is caused by binary-interaction processes. In such cases, the (candidate) 
LBV stars might be unresolved binaries or merger products of two stars, while the nearby stars might represent the tertiary stars 
dynamically scattered in wide orbits (cf. Reipurth \& Mikkola 2012) or former (unbound) members of dissolved triple systems (cf. 
Gvaramadze \& Menten 2012). BD+60$\degr$\,2668B might be such a tertiary star.

In summary, the spectral analysis of BD+60$\degr$\,2668A\&B indicates that both stars are either still on the main sequence
or only recently left it, and that the surface nitrogen abundance of BD+60$\degr$\,2668A is enhanced by a factor of several. Also, it 
appears that BD+60$\degr$\,2668A is a factor of two younger than BD+60$\degr$\,2668B despite the strong indications that both stars are 
members of the same stellar system and, therefore, most likely were formed simultaneously. The age discrepancy could be naturally 
understood if BD+60$\degr$\,2668A is a rejuvenated product of binary interaction, in which case the enhanced nitrogen abundance on 
the surface of the star is a direct consequence of this interaction. The binary interaction scenario also provides a framework for 
understanding the origin of the circumstellar nebula, which hardly can be produced by a single unevolved star.

\subsection{WS29 and ALS\,19653}
\label{sec:ws29}

For ALS\,19653 we derive $M_V\approx-5.5$ mag and $\log(L/\lsun)\approx5.1$, which along with the effective temperature imply 
an initial mass and age of ALS\,19653 of, respectively, $\approx20 \, \msun$ and $\approx7-8$ Myr, meaning that this star is either 
still on the main sequence or just left it (e.g. Brott et al. 2011).

The chemical abundances derived for ALS\,19653 (see Table\,\ref{tab:abu}) support the inference that this star is relatively unevolved,
and agree with the Brott et al.'s models if the initial rotational velocity of this star was $\sim200$~\kms.
Note that the uncertainties in the measurements allow the possibility that the surface nitrogen abundance is enhanced by factors of 2 
or 3 in comparison with the CAS or the initial N abundance adopted in Brott et al. (2011), respectively. This enhancement, if real, 
could be caused not only by rotationally-induced mixing, but also by binary interaction, e.g., because of accretion of CNO-processed 
material from the companion star or mixing caused by merger of the binary components (e.g. Langer 2012). 

The detection of two nested shells around ALS\,19653 indicates that this star has experienced at least two episodes of enhanced
mass loss\footnote{Similar two-shell structure was also found around the O6.5f?p (Walborn et al. 2010) star HD\,148937 
(Leitherer \& Chavarria-K. 1987). It is believed that the strong magnetic field of this star, 
its rapid rotation and the origin of the nitrogen-rich circumstellar nebula are the result of merger of two stars (Langer 2012).}. 
At the adopted distance to ALS\,19653, the characteristic radius of the second (outer) shell is $\approx$1.2 
pc, which along with the expansion velocity of this shell of $\sim100$~\kms gives its kinematic age of $\sim10^4$ yr.
Similarly, assuming that the optical shell of WS29 is expanding with a velocity of $20-30$~\kms (cf. Fig.\,\ref{fig:ha}) and given
its characteristic radius of $\approx0.2$ pc, one finds that this shell was formed shortly after the outer shell. WS29 is especially 
curious because it consists of an optical circular shell with protrusions in the northwest and southeast directions, and an inner 
elongated IR nebula, stretched in the direction perpendicular to the axis defined by the protrusions. Below we discuss the possible 
origin of WS29.

The changes in the H$\alpha$ line profile suggest that ALS\,19653 is surrounded by a disc-like 
structure viewed nearly edge-on (see Section\,\ref{sec:mod}), while the polarization angle of the stellar light of 121\degr 
(see Section\,\ref{sec:two}) implies that this structure is oriented in the same direction as the IR nebula. We therefore interpret 
this nebula as a flattened outflow or ejecta from ALS\,19653 viewed nearly edge-on. Since it is unlikely that a single unevolved 
star can produce a circumstellar nebula it is reasonable to assume that its formation is due to a binary 
interaction process, e.g., because of merger of two stars or common-envelope evolution. Correspondingly, the spin axis of the merger 
product or the angular momentum of the post-common-envelope binary orbit should coincide with the line connecting the protrusions 
in the optical shell of WS29 (PA$\approx$120\degr). If ALS\,19653 is indeed the result of merger of two stars, then the inclination angle 
of its rotational axis to our line of sight should be close to 90$\degr$, while its equatorial rotational velocity should be 
almost equal to the projected rotational velocity of 28~\kms. The H$\alpha$ line variability in ALS\,19653 could be understood if the 
density structure of the (inner) part of the flattened circumstellar nebula is non-axisymmetric, e.g. 
because of the tidal force from the companion star (cf. Okazaki et al. 2002; Oktariani \& Okazaki 2009). 
The possible binary nature of ALS\,19653 (see Section\,\ref{sec:mod}) makes this explanation plausible. It should be noted, however, 
that binarity of ALS\,19653 does not exclude the possibility that this star is a merger product because originally it might have been a 
triple system (cf. Pasquali et al. 2000; Schneider et al. 2016).

The strong differential rotation originating in the course of merger of a binary system or common envelope evolution might be 
responsible for generation of a strong large-scale magnetic field in the newly formed single star (Langer 2012; Wickramasinghe, 
Tout \& Ferrario 2014; Schneider et al. 2016) or in the ejected envelope (Reg\'os \& Tout 1995; Tout \& Reg\"os 2003), 
respectively. This magnetic field can in turn effectively spin down the merger product through dynamical mass-loss (e.g. Langer
2012) and may affect 
(or even determine) the geometry of the ejected material (Chevalier \& Luo 1994; R\'o\.zyczka \& Franco 1996; Garcia-Segura et 
al. 1999; Nordhaus \& Blackman 2006). Correspondingly, the low rotational velocity derived for ALS\,19653 might be the direct 
result of magnetic braking of this star, while the origin of protrusions in the northwest and southeast directions might be caused 
by deflection of the ejected material towards the polar directions by the tension of the toroidal magnetic field. The 
gravity-darkening of the initially fast-rotating merger remnant may also contribute to the origin of the polar protrusions.

Finally, we note that the distribution of the H$\alpha$ radial velocity along the slit with PA=33$\degr$ (i.e. along the slit 
aligned with the IR nebula) suggests that the outer parts of the equatorial ejecta might be rotating around the axis 
defined by the orientation of the polar protrusions. If real, this rotation would imply very efficient transfer of 
angular momentum from the merger remnant (possibly rotating at break-up velocity at the equator) or from the shrinking 
binary system to a tiny fraction of the equatorial ejecta by means of magnetic torque. Follow-up study of the H$\alpha$ velocity 
field in WS29 with the multi-unit spectroscopic explorer (MUSE; Bacon et al. 2010) could potentially clarify this issue.

\section{Summary}
\label{sec:sum}

We discovered two mid-infrared nebulae using data from the {\it WISE} survey, and identified their central stars with
the early B stars BD+60$\degr$\,2668 and ALS\,19653. Our spectroscopic observations of BD+60$\degr$\,2668 (composed of two
components, A and B, separated by $\approx3$ arcsec) and ALS\,19653 confirmed that they are massive stars. We analysed spectra
of these stars with the stellar atmosphere code {\sc fastwind} and found that all three stars are either on the main sequence
or just left it. The obtained results provide further support to claims that massive stars can produce circumstellar nebulae
during the early stages of evolution. Since the origin of such nebulae hardly can be understood within the framework of
single star evolution, and because the majority of massive stars form in binary or multiple systems and many of them
undergo binary interaction during the main sequence stage, we suggested that the formation of both nebulae is related to the
binary nature of their central stars. We found that the surface nitrogen abundance of BD+60$\degr$\,2668A is enhanced by a
factor of several with respect to the solar value and that the apparent age of this star is a factor of 1.5 younger than that
of BD+60$\degr$\,2668B. Proceeding from this, we suggested that component\,A is a rejuvenated product of binary merger and that
the mass lose from just this star is responsible for the origin of the nebula around BD+60$\degr$\,2668. We also found that
ALS\,19653 shows significant radial velocity variations, meaning that it is a close binary system. This and the presence of a
disc-like structure around ALS\,19653 (as suggested by polarimetric, infrrared and radio observations, and variable
profile of the H$\alpha$ emission line in the spectrum of this star) suggest that the origin of the associated nebula might
also be caused by binary interaction.

\acknowledgments

This work is based on observations collected with the Southern African Large Telescope (SALT), programmes 2016-1-SCI-012 and 
2017-1-SCI-006, and at the Centro Astron\'omico Hispano Alem\'an (CAHA), operated jointly by the Max-Planck Institut f\"ur 
Astronomie and the Instituto de Astrofisica de Andalucia (CSIC), programme H12-3.5-013, and supported by the Russian Foundation 
for Basic Research grant 16-02-00148. AYK acknowledges support from the National Research Foundation (NRF) of South Africa. 
EKG gratefully acknowledges funding by the Sonderforschungsbereich ``The Milky Way System'' (SFB\,881, especially subproject A5) 
of the German Research Foundation (DFG).
This work has made use of data products from the Wide-field Infrared Survey Explorer, which is a joint project of the University 
of California, Los Angeles, and the Jet Propulsion Laboratory/California Institute of Technology, funded by the National 
Aeronautics and Space Administration, the SIMBAD data base and the VizieR catalogue access tool, both operated at CDS, 
Strasbourg, France. This work also has made use of data from the European Space Agency (ESA) mission
{\it Gaia} (https://www.cosmos.esa.int/gaia), processed by the {\it Gaia}
Data Processing and Analysis Consortium (DPAC,
https://www.cosmos.esa.int/web/gaia/dpac/consortium). Funding for the DPAC
has been provided by national institutions, in particular the institutions
participating in the {\it Gaia} Multilateral Agreement.

\software{{\sc fastwind} (Santolaya-Rey, Puls \& Herrero 1997; Puls et al. 2005; Rivero Gonz{\'a}lez et al. 2012), 
{\sc iraf} (Tody 1986, Tody 1993), {\sc iacob-broad} (Sim\'on-Diaz \& Herrero 2007, 2014), {\sc ulyss} (Koleva et al. 2009)}

\end{document}